\documentclass[prl,twocolumn,showpacs,amsmath,amssymb,citeautoscript,nobibnotes,preprintnumbers]{revtex4-1}

\usepackage{graphicx,times}
\usepackage{array}
\usepackage{bm}
\usepackage{amsfonts}
\usepackage{graphics}
\usepackage{mathrsfs}
\usepackage{amssymb}

\newcommand{\beq}{\begin{equation}}
\newcommand{\eeq}{\end{equation}}
\newcommand{\beqa}{\begin{eqnarray}}
\newcommand{\eeqa}{\end{eqnarray}}

\newcommand{\chib}{{\overline{\chi}}}

\newcommand{\mcall}{{\mathbb{L}}}
\newcommand{\mcalb}{{\mathbb{B}}}
\newcommand{\mcalf}{{\mathbb{F}}}

\newcommand\comment[1]{}

\begin{document}

\title{Quantum critical behavior in three dimensional lattice Gross-Neveu models}

\author{Shailesh Chandrasekharan}
\affiliation{Department of Physics, Duke University, Durham, NC 27708, USA}
\author{Anyi Li}
\affiliation{Institute for Nuclear Theory, University of Washington, Seattle, 98195, USA}
\preprint{INT-PUB-13-016}

\keywords{Sign Problem, Gross-Neveu models, Chiral Symmetry}
\begin{abstract}
We study quantum critical behavior in three dimensional lattice Gross-Neveu models containing two massless Dirac fermions. We focus on two models with $SU(2)$ flavor symmetry and either a $Z_2$ or a $U(1)$ chiral symmetry. Both models could not be studied earlier due to sign problems. We use the fermion bag approach which is free of sign problems and compute critical exponents at the phase transitions. We estimate $\nu = 0.83(1)$, $\eta = 0.62(1)$, $\eta_\psi = 0.38(1)$ in the $Z_2$ and $\nu = 0.849(8)$, $\eta = 0.633(8)$, $\eta_\psi = 0.373(3)$ in the $U(1)$ model.
\end{abstract}
\pacs{71.10.Fd, 02.70.Ss,11.30.Rd,05.30.Rt, 05.50.+q, 03.70.+k}
\maketitle

The presence of massless Dirac fermions at low energies in graphene, has created much excitement over the past decade \cite{RevModPhys.81.109,RevModPhys.84.1067}. By increasing the interaction strength between the electrons experimentally, an energy gap can be opened and fermions can become massive \cite{Zhou2007,PhysRevLett.101.146805}. Such quantum phase transitions, between a massless (semi-metal) and a massive (insulator) phase, are well known in particle physics. The possibility of studying them in a laboratory has ignited interest in the subject recently \cite{PhysRevB.72.085123,nature.464.08942,Sorella2012}. Renormalization group arguments suggest that non-relativistic effects and long range interactions could be irrelevant \cite{PhysRevB.75.235423}, and the transition could belong to the universality class of similar phase transitions in three dimensional relativistic four-fermion field theories with two massless Dirac fermions \cite{PhysRevLett.97.146401,Mesterhazy:2012ei}. While Monte Carlo calculations of the critical exponents in models of graphene have emerged recently, the results are neither consistent with each other \cite{PhysRevB.81.125105,PhysRevB.79.165425,PhysRevB.79.241405} nor do they match theoretical predictions \cite{PhysRevLett.97.146401}. Our current understanding of related four-fermion field theories is also quite limited. Compared to the precision with which three dimensional Ising and XY models have been studied \cite{PhysRevB.59.11471,PhysRevB.63.214503}, critical exponents in models with similar symmetry breaking patterns but in the presence of two massless Dirac fermions at the critical point remain largely unknown. As we explain below, some existing results are even puzzling. In this work we report new results in models that could not be studied earlier due to sign problems. Our results clarify some puzzles and help understand the subject better.

Relativistic four-fermion models have a long history and are usually studied in the presence of either scalar interactions (Gross-Neveu models) or vector interactions (Thirring models) \cite{Klimenko:1987gi,Rosenstein:1988pt,ZinnJustin:1991yn,Rosenstein:1990nm,PhysRevD.43.3516,Hands:1994kb}. Their lattice formulations using staggered fermions are popular, but due to fermion doubling one flavor of staggered fermions in three dimensions produces two flavors ($N_f=2)$ of Dirac fermions \cite{AnnPhys.224.29,DelDebbio:1995zc}. Symmetries of the microscopic models play an important role in determining the universality class of phase transitions. Gross-Neveu models with a variety of symmetries have been studied using large $N_f$ expansions \cite{Vasiliev:1992wr,Gracey:1993kc}, $\epsilon$-expansions \cite{Rosenstein:1993zf}, renormalization-group (RG) flow methods \cite{Berges:2000ew,PhysRevB.66.205111,Scherer:2012tc}, and lattice Monte Carlo calculations \cite{Karkkainen:1993ef,PhysRevD.53.4616,Christofi:2006zt}. Although much has been understood, there remain puzzles. For example, the critical exponents in the continuum Gross-Neveu model with a $U(4)\times Z_2$ symmetry computed with the RG-flow method \cite{PhysRevB.66.205111}, match those calculated with lattice Monte Carlo methods in a model with an $SU(2)\times Z_2$ symmetry \cite{Karkkainen:1993ef}. Both models contain two flavors of Dirac fermions and calculations give $\nu \approx 1.0$ and $\eta \approx 0.75$. Why do models with two different symmetries lead to the same critical behavior? Are symmetries dynamically enhanced in the lattice model at the critical point? On the other hand, is it possible that the results of Ref. \onlinecite{Karkkainen:1993ef} are incorrect since sign problems were ignored \cite{Chandrasekharan:2012va,Chandrasekharan:2012fk}? Here we show that another lattice model with the same symmetries give different critical exponents. Another puzzle concerns a comparison between calculations of critical exponents in the continuum Thirring model with $U(4)$ symmetry computed recently using the RG-flow method \cite{,Janssen:2012pq}, and those in a lattice Thirring model with an $SU(2) \times U(1)$ symmetry obtained with Monte Carlo calculations that do not suffer from sign problems \cite{Debbio:1997dv,Barbour:1998yc,Chandrasekharan:2011mn}. While both models contain two flavors of Dirac fermions, in the continuum one finds $\nu \approx 2.4$ and $\eta \approx 1.4$ while in the lattice one finds $\nu \approx 0.85$ and $\eta \approx 0.65$. Although this disagreement can be attributed to the difference in the symmetries, it does raise the question if lattice calculations have uncovered a new universality class. Here we show that lattice Gross-Neveu models defined in \cite{AnnPhys.224.29} and lattice Thirring models defined in \cite{Debbio:1997dv} have the same symmetries and critical exponents.

\begin{figure}[t]
\hbox{
\includegraphics[width=1.1in]{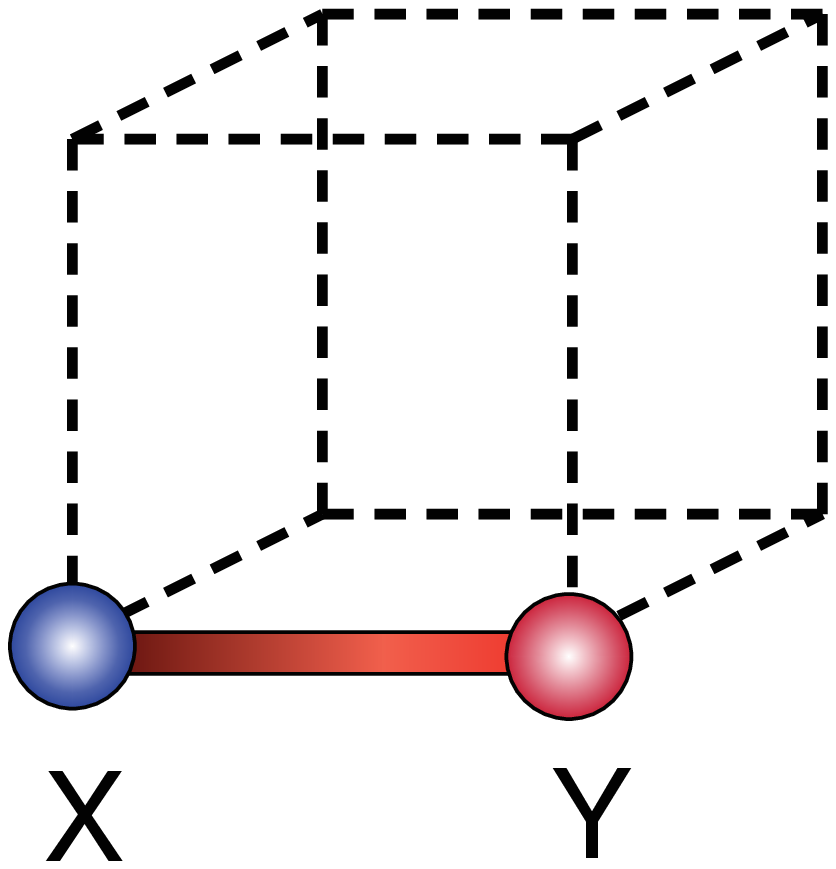}
\includegraphics[width=1.1in]{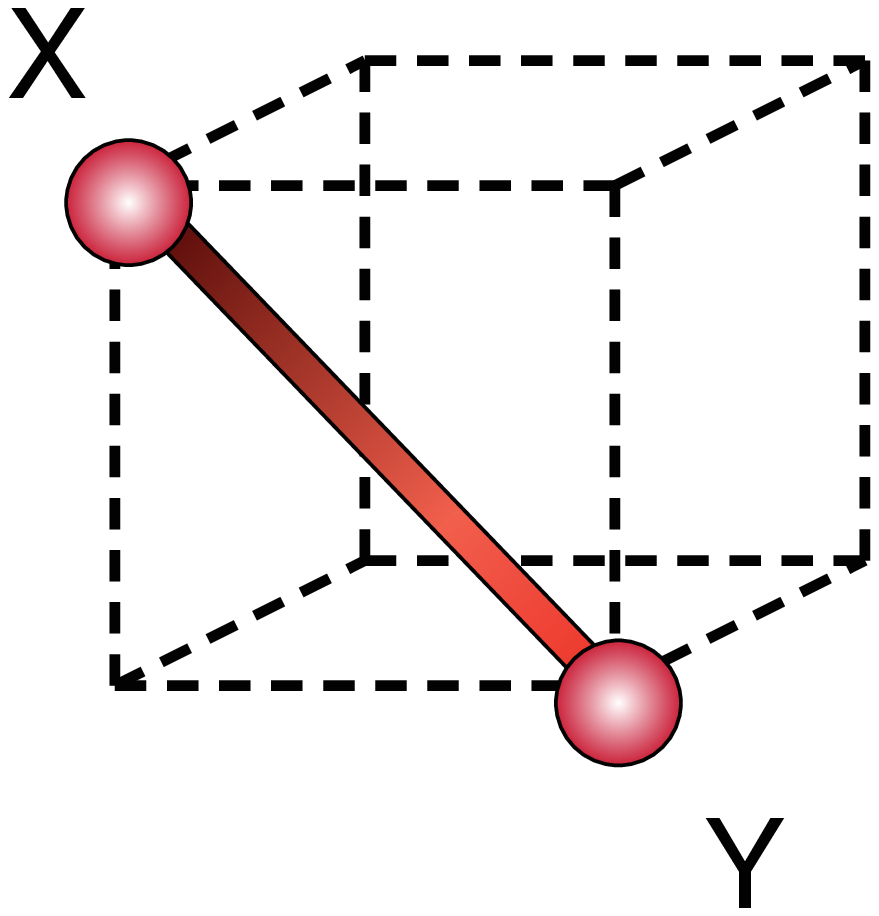}
\includegraphics[width=1.1in]{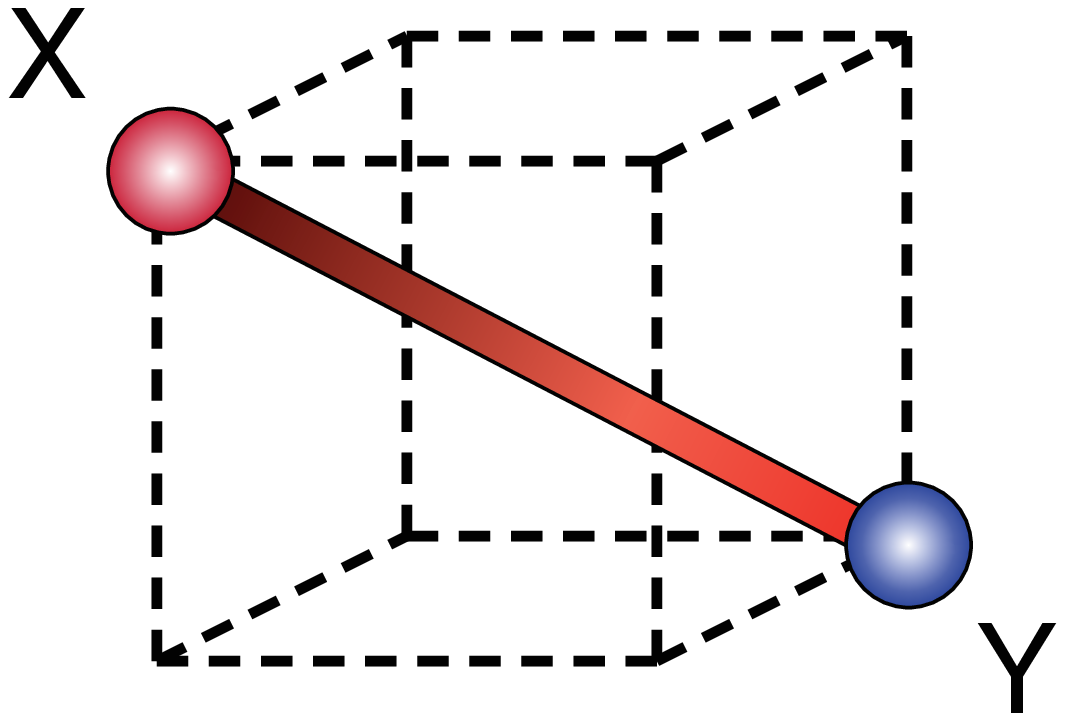}
}
\caption{\label{fig1} A pictorial representation of the bond couplings $U_\mcall$ (left), $U_\mcalf$ (center) and $U_\mcalb$ (right) discussed in the text. Each bond refers to the four-fermion interaction term of the form $\chib_x\chi_x\ \chib_y\chi_y$.}
\end{figure}

Lattice Gross-Neveu models with one flavor of staggered fermions cannot be solved reliably  in the traditional approach due to sign problems \cite{Chandrasekharan:2012va,Chandrasekharan:2012fk}. The fermion bag approach is an alternative method which is free of sign problems and allows one to perform computations in these models reliably for the first time \cite{Chandrasekharan:2009wc,Chandrasekharan:2013rpa,Chandrasekharan:2011mn}. We use this new method to compute critical exponents in two types of lattice Gross-Neveu models invariant under either a $Z_2$ or a $U(1)$ chiral symmetry. The models also have an additional $SU(2)$ flavor symmetry which was appreciated only recently. Since they naturally describe two flavors of Dirac fermions in the critical region, these models have many properties similar to graphene including symmetries. They were formulated originally with auxiliary fields that live at the center of cubes and couple to fermions on the corners \cite{AnnPhys.224.29}. After integrating over the auxiliary fields we obtain four-fermion models that couple fermion fields within a hypercube. Their action can be written as
\beq
S = \sum_{x,y}\chib(x) \ D_{xy} \ \chi(y) \ -\  \sum_{\langle xy \rangle} 
U_{\langle xy\rangle} \chib_x\chi_x\ \chib_y\chi_y
\label{model}
\eeq
where $\chib(x), \chi(x)$ denote two Grassmann valued fermion fields at the lattice site $x$ and $D$ is the free massless staggered fermion matrix defined by
\beq
D_{xy}  =  
\frac{1}{2}\sum_{\alpha}\ \eta_{x,\alpha}\ \left[\delta_{x+\alpha,y} \ -\  \delta_{x,y+\alpha}\right],
\label{staggDirac}
\eeq
in which $\alpha$ labels the three directions and $\eta_{x, \alpha}= e^{(i \pi \zeta_a \cdot x)}, \zeta_1=(0,0,0)$, $\zeta_2=(1,0,0)$, $\zeta_3=(1,1,0)$ are the staggered fermion phase factors \cite{Sharatchandra:1981si}. The four-fermion interaction term involves the sum over three types bonds denoted by $\langle xy \rangle$ (see Fig.~\ref{fig1}): (1) link bonds $\mcall$ (where $x,y$ are nearest neighbor sites), (2) face bonds $\mcalf$ (where $x,y$ are sites diagonally across faces of squares), (3) body bonds $\mcalb$ (where $x,y$ are sites diagonally across the bodies of cubes). 

In a general lattice four-fermion model the three couplings $U_\mcall$, $U_\mcalf$ and $U_\mcalb$ will be arbitrary. However, in our study they are constrained since the action (\ref{model}) is obtained by integrating over auxiliary fields from a model that contains a single coupling. In the Gross-Neveu model with $Z_2$ chiral symmetry, we find $U_\mcall = 2 U_\mcalf = 4 U_\mcalb \equiv U$, while with $U(1)$ chiral symmetry we find $U_\mcall  = 4 U_\mcalb \equiv U, U_\mcalf = 0$ \cite{Chandrasekharan:2012va}. In other words, face diagonal bonds break the $U(1)$ symmetry to $Z_2$. In addition to chiral symmetries, models with action (\ref{model}) have an $SU(2)$ flavor symmetry which is hidden in the auxiliary field approach and was not appreciated earlier \cite{Catterall:2011ab}. Indeed, when $U_\mcalf = 0$ it is easy to verify that the action (\ref{model}) is invariant under the following $SU(2) \times U(1)$ symmetry,
\beq
\left(\begin{array}{c} \chi_e \cr \chib_e \end{array}\right) 
\rightarrow \mathrm{e}^{i\theta} V \left(\begin{array}{c} 
\chi_e \cr \chib_e \end{array}\right), \ \ 
\left(\begin{array}{cc} \chib_o & \chi_o \end{array}\right) \rightarrow 
\left(\begin{array}{cc} \chib_o & \chi_o \end{array}\right) 
V^\dagger \mathrm{e}^{-i\theta},
\eeq
where the subscripts $e$ and $o$ refer to even and odd sites and $V$ is an $SU(2)$ matrix. When $U_\mcalf \neq 0$ the symmetry is restricted to $\theta = \pi/2$ and the action is invariant only under an $SU(2) \times Z_2$ symmetry.

\begin{figure}[t]
\includegraphics[width=3.3in]{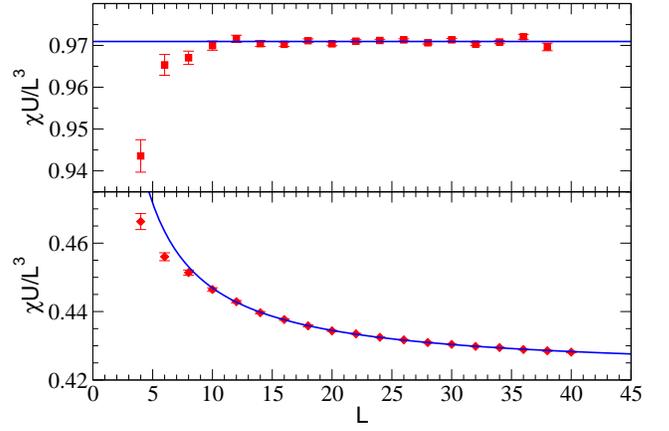}
\caption{Plot of the chiral susceptibility at $U=\infty$ for the $Z_2$ (top) and $U(1)$ (bottom) models. The solid curve in the top graph is a fit to the constant for $L \geq 16$, while in the bottom graph it is a fit to the finite size scaling form (\ref{chpt}) for $L \geq 10$ obtained from chiral perturbation theory. \label{fig2}}
\end{figure}

Since four-fermion couplings are perturbatively irrelevant in three dimensions, models with action (\ref{model}) have a massless fermion phase at small couplings $U$. As the coupling increases, a second order phase transition to a massive fermion phase accompanied by spontaneous breaking of chiral symmetries occurs at a critical coupling $U_c$. Our goal is to study the critical exponents at this transition. However, before focusing on the transition region, it is useful to understand qualitatively the physics of the massive phase at large $U$. There is an important difference between spontaneous breaking of $Z_2$ and $U(1)$ symmetries; the former does not produce massless Goldstone bosons while the latter does. It is important to distinguish this feature in our results. For this purpose we have computed the chiral condensate susceptibility,
\beq
\chi = \frac{1}{L^3}\sum_{x,y}\langle\chib_x\chi_x\chib_y\chi_y\rangle,
\label{csus}
\eeq
as a function of the lattice size $L$ at $U=\infty$. At infinite coupling our models can be mapped into a statistical model of closed packed dimers and can be updated efficiently using worm algorithms \cite{Prokof'ev:2001zz}. Results obtained are shown in Fig.~\ref{fig2}. As expected, finite size effects are enhanced in the $U(1)$ invariant model due to the presence of massless Goldstone bosons. Results for $L\geq 10$ fit well to the leading order chiral perturbation theory form \cite{Hasenfratz:1989pk}
\beq
\chi/L^3 = \frac{\Sigma^2}{2}\  \big(1+0.224/(\rho_s L) \big),
\label{chpt}
\eeq
with  $\Sigma^2 = 0.844(1)$, $\rho_s=0.381(3)$ and $\chi^2/d.o.f = 0.4$. In contrast, the $Z_2$ model shows very small finite size effects which indicates the absence of massless modes, and the data for $L \geq 16 $ fits the constant $0.971(1)$ with a $\chi^2/d.o.f = 1.7$. 

\begin{table*}[t]
\begin{center}
\begin{tabular}{c c c c c c c c c c c c c c c}
\hline\hline
$U_c$ & $\nu$ & $\eta$ & $\eta_\psi$ & $f_0$ & $f_1$ &$f_2$& $f_3$ & $f_4$  & $p_0$& $p_1$ &  $p_2$ & $p_3$ & $p_4$ & $\chi^2$/d.o.f\\
\hline
0.0893(1) & 0.83(1) & 0.62(1) & 0.38(1) & 2.54(7) & 9.33(5) & 27.3(3) & 55.3(1) &48.67(3)  & 34.4(1) & -18.2(7) & -51.2(6) & 7.4(4) & 259.2(10) & 1.8\\
\hline
0.1560(4) & 0.82(2) & 0.62(2) & 0.37(1) & 0.13(1) & 0.09(1) & 0.02(1) & 0.004(1) &0.02(1)  & 34.0(1) & -4.5(3) & -1.4(3) & -1.8(8) & -0.5(2) & 0.88\\
\hline\hline
\end{tabular} 
\caption{Results of the combined fit of data in the critical region to Eqs.~(\ref{eq:scaling}) in the $Z_2$ invariant model (top row) and $U(1)$ invariant model (bottom row).}
\label{tabfit}
\end{center}
\end{table*}

\begin{figure*}[t]
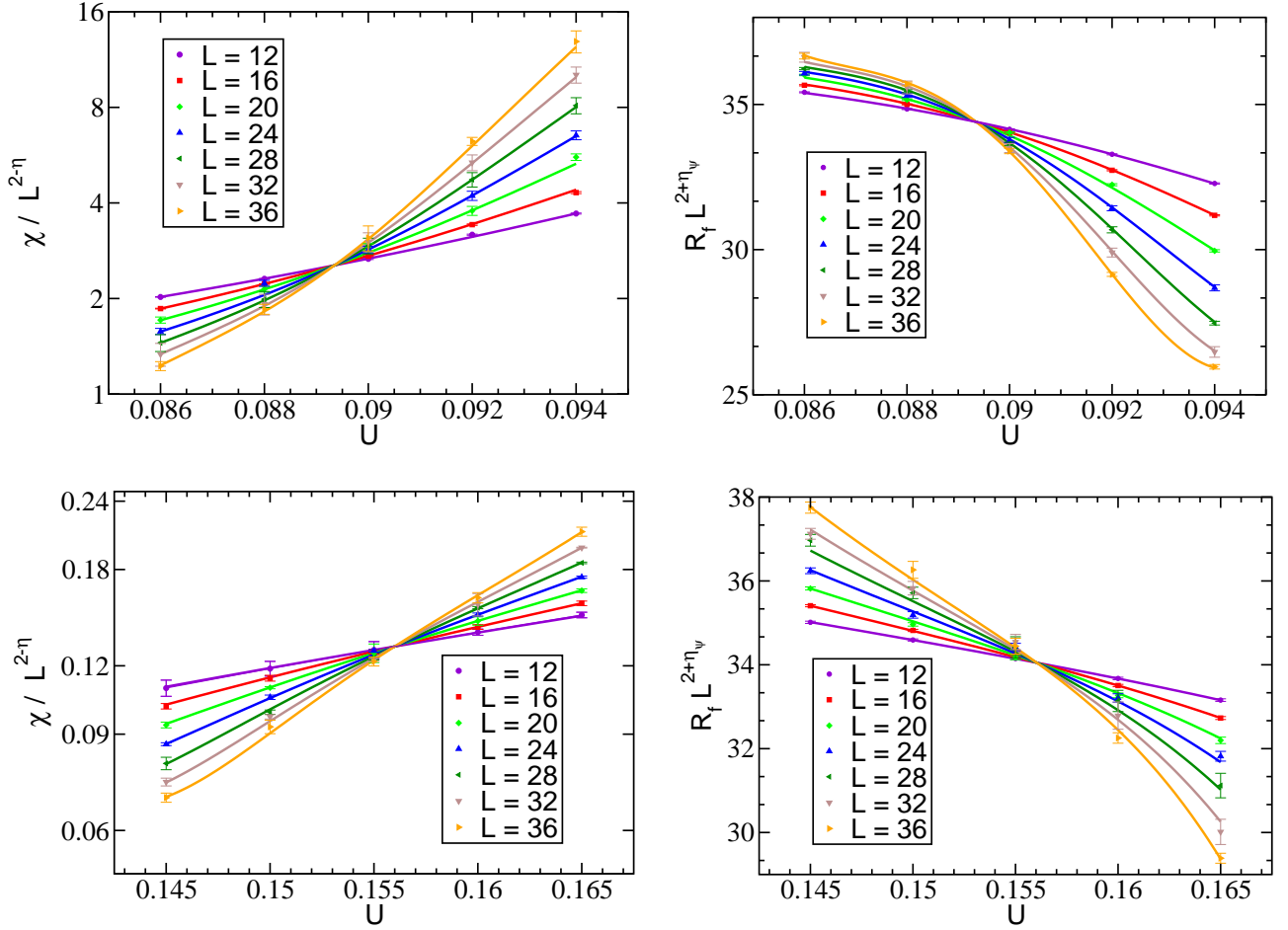

\begin{center}
\hbox{
\includegraphics[width=3.3in]{susc_z2.eps}
\hskip0.2in
\includegraphics[width=3.15in]{fratio_z2.eps}
}
\vskip0.2in
\hbox{
\hbox{
\includegraphics[width=3.3in]{susc_u1.eps}
\hskip0.2in
\includegraphics[width=3.15in]{fratio_u1.eps}
}
}
\caption{Plots of $\chi / L^{2-\eta}$ and $R_f L^{2+\eta_\psi}$ as a function of $U$ for $L$ from $12$ to $36$. The solid lines show the combined fit which gives $U_c = 0.0893(1), \nu = 0.83(1), \eta=0.62(1)$ and $\eta_\psi=0.38(1)$ in the $Z_2$ case (top row) and $U_c = 0.1558(4), \nu = 0.82(2), \eta=0.63(2)$, $\eta_\psi=0.37(1)$ in the $U(1)$ case (bottom row).
}
\label{fig:scaling}
\end{center}
\end{figure*}

In order to uncover the properties of the quantum critical point we focus on the chiral susceptibility (\ref{csus}) and the fermion correlation function ratio
\begin{subequations}
\beqa
R_f&=&C_F(L/2-1)/C_F(1), \\
C_F(d) &=& \frac{1}{3}\sum_{\alpha = 1}^3 \langle\chi_x\ \ \chib_{x+d\hat{\alpha}}\rangle
\eeqa
\end{subequations}
where $x$ is the origin or any translation of it by a multiple of two lattice spacings in each direction, and $\hat{\alpha}$ is a unit vector along each of the three directions. Since fermions are exactly massless, in the vicinity of $U_c$ we expect $\chi$ and $R_f$ to satisfy the following universal finite size scaling relations:
\begin{subequations}
\label{eq:scaling}
\beqa
\chi / L^{2-\eta} &=& \sum_{k=0}^4 f_k \left[(U-U_c) L^{\frac{1}{\nu}}\right]^k, \\
R_f L^{2+\eta_\psi} &=& \sum_{k=0}^4 p_k \left[(U-U_c) L^{\frac{1}{\nu}}\right]^k, 
\eeqa
\end{subequations}
where we have kept the first five terms in the Taylor series of the corresponding analytic functions. In order to compute the critical exponents $\eta$, $\nu$ and $\eta_\psi$ we perform a single combined fit of the data in the critical region to Eqs.~(\ref{eq:scaling}) with fourteen parameters. For the $Z_2$ invariant model the combined fit of the data using lattice sizes ranging from $12^3$ to $36^3$ gives $\nu = 0.83(1)$, $\eta=0.62(1)$, $\eta_\psi=0.38(1)$  and $U_c = 0.0893(1)$ with a  $\chi^2/d.o.f.=1.8$. For the U(1) Gross-Neveu model, a similar combined fit in the same range of lattice sizes gives $\nu = 0.82(2)$, $\eta=0.62(2)$, $\eta_\psi=0.37(1)$, $U_c = 0.1560(4)$ with a  $\chi^2/d.o.f.=0.88$. Plots of our data along with the fits are shown in Fig.~\ref{fig:scaling}. The complete list of the fourteen fit parameters are listed in Tab.~\ref{tabfit}. From the results above, it seems like the critical exponents do not change much
when chiral symmetries change from $Z_2$ to $U(1)$; the differences are small and lie within error bars.

The critical exponents in the $SU(2)\times U(1)$ symmetric lattice Gross-Neveu model obtained here, are also consistent with the exponents in the lattice Thirring model, which also has an action of the form (\ref{model}) except that $U_\mcall = U, U_\mcalf = U_\mcalb = 0$ \cite{Chandrasekharan:2011mn}.  This is reassuring since the two models have the same lattice symmetries. Thus, calling one as the lattice Gross-Neveu model and the other as the lattice Thirring model is just a matter of taste. We can also study other $SU(2)\times U(1)$ invariant models by choosing a different set of couplings. Recently, the model with $U_\mcall = U_\mcalb = U, U_\mcalf = 0$ was also studied and the critical exponents were again found to be similar \cite{Li:2012dra}. Thus, it is tempting to combine all data from the three different studies and perform a single combined fit to extract a more accurate set of critical exponents. Using such a fit we estimate the critical exponents in the $SU(2)\times U(1)$ symmetric lattice models to be $\nu = 0.849(8)$, $\eta = 0.633(8)$ and $\eta_\psi = 0.373(3)$.

Interestingly, the model studied in Ref.~\onlinecite{Karkkainen:1993ef} is also an $SU(2) \times Z_2$ symmetric Gross-Neveu model. It is slightly different from the model studied here since the auxiliary fields in the defining model live on sites instead of centers of hypercubes. Integration over the auxiliary fields, which couple fermions on the six neighboring sites, produces four-fermion couplings of the form given in the action (\ref{model}) with $U_\mcall = U_\mcalb = 0,U_\mcalf = U$. However, in addition there is a non-zero next-to-nearest-neighbor four-fermion coupling along each direction, which is not present in our work. Since no lattice symmetries change, it seems very unlikely that these differences change the universality class of the phase transition. Hence, we believe the critical exponents of the model studied in Ref. \onlinecite{Karkkainen:1993ef} should have been identical to our studies here. Unfortunately, this is not the case and we think that ignoring the sign problem in the auxiliary field approach could have distorted the results. It would be useful to repeat the calculation with the fermion bag approach.

In this work we have been able to accurately compute the critical exponents at phase transitions in a class of $SU(2)\times Z_2$ and $SU(2) \times U(1)$ symmetric four-fermion models involving two massless Dirac fermions in three dimensions. The critical exponents of the two models match within errors and a more accurate calculation is necessary to distinguish between them. Since the symmetries are different, we do not see any reason for the two exponents to be the same, however we are unable to rule out this possibility at the moment. As far as we can tell these critical exponents have not been verified in continuum field theory by the recently developed RG-flow method. However, we note that the $\epsilon$-expansion to second order in a Gross-Neveu model does agree with our results for the exponents $\nu$ and $\eta$, but not for $\eta_\psi$ \cite{Rosenstein:1993zf}. Finally, given many similarity between graphene and staggered fermions, it would be interesting if the critical behavior in graphene falls in one of the universality classes studied here.

\section*{Acknowledgments}
It is a pleasure to thank H.~Gies, S.~Hands, L.~Janssen, B.~Rosenstein and C.~Strouthos for their time and patience in answering many of our questions and helping us begin to understand this complex subject. This work was supported in part by the Department of Energy grants DE-FG02-05ER41368 and DE-FG02-00ER41132.

\bibliography{fourfermi}
\end{document}